\documentclass[aps,prl,twocolumn,preprintnumbers,amsmath,amssymb,floatfix,superscriptaddress]{revtex4}
\usepackage[dvips]{graphicx}

\DeclareGraphicsExtensions{.eps,.ps}

\def\s#1{_{\rm #1} }

\def\cI{ {\cal I } }
\def\cA{ {\cal A } }
\def\e{ {\rm e } }
\def\bea{\begin{eqnarray}}
\def\eea{\end{eqnarray}}
\def \be{\begin{equation}}
\def \ee{\end{equation}}

\def\trans{\textit{trans}\,}
\def\cis{\textit{cis}\,}
\usepackage{hyperref}
\begin{document}

\title{The dynamics of non-linear optical absorption}
\author{D. Corbett}
\affiliation{Cavendish Laboratory, JJ Thomson Avenue, Cambridge CB3
0HE, United Kingdom}
\author{ C.L. van Oosten}
\affiliation{Department of Chemistry and Chemical Engineering,
Technische Universiteit Eindhoven, Postbus 513, 5600 MB, Eindhoven,
The Netherlands}
\author{M. Warner}
\affiliation{Cavendish Laboratory, JJ Thomson Avenue, Cambridge CB3
0HE, United Kingdom}

\date{\today}
\begin{abstract}
On traversing materials with absorbing dyes, weak optical beams
develop a Beer (exponential) profile, while intense beams develop a
spatially initially linear and then finally an exponential profile.
This anomalous, deep penetration due to photo-bleaching of surface
layers is important for heavy dye-loading and intense beams, for
instance in  photo-actuation. We address the problem of the
evolution in time from initial Beer's Law to deeply penetrating
optical profiles in dyes.  Our solution of the coupled, non-linear,
partial differential equations governing the spatio-temporal decay
of the Poynting flux and the non-linear
 dynamics of the \textit{trans-cis} conversion is applicable to
general systems of photo-active  molecules under intense
irradiation, for instance in biology, in spectroscopy and in
opto-mechanical devices.
 \end{abstract}
 \maketitle

Light absorption can optically switch the nematic phase of liquid
crystals to the disordered, isotropic state: rodlike dye molecules
transform from their linear \textit{trans} to their bent
\textit{cis} isomers and thereby disrupt the orientational order of
their nematic hosts.  Although an interesting optical effect in its
own right, when such a host is part of a nematic elastomer, then the
solid responds with stresses \cite{Cviklinski:03} or huge (100s\%)
optically-induced strains \cite{Pallfy:04,Harris:05}. Optical
actuation offers advantages over electrical, thermal and solvent
mediated response, being of large amplitude, easily reversible, sensitive to polarisation and susceptible to remote application.

Two mysteries attend  photoisomerisation, seen for instance in actuation.
For linear absorptive processes at low light intensities (and
therefore exponential profiles with depth, i.e. Beer's Law), and for
mechanical contraction proportional to \textit{trans-cis}
conversion, one can show \cite{Mahadevan04} that optimal bend of a
photo-cantilever obtains at $w/d \sim 3$ where $w$ is the cantilever
thickness and $d$ the  penetration depth in Beer's Law $\cI =
I(x)/I(0) = \e^{-x/d}$ with $I(x)$ the Poynting flux at depth $x$ in
the cantilever.  For $d\ll w$ only a thin skin of material near the
front face contracts and the remaining material resists bend.  This limit of
little penetration is that of high dye loading and in practice is
often employed~\cite{vanOosten:07b}. Non-linear absorption at high light
intensities, where Beer's law gives way to a linear and more
penetrating profile for $\cI(x)$ \cite{Statman:03}, has been
suggested as a mechanism \cite{Corbett_PRL:07,Corbett_PRE:08} to
explain the bending of cantilevers with $d\ll w$.  The second mystery concerns the dynamical response, which is often initially slow and then proceeds quickly  \cite{Serra:08}, especially in heavily dye-loaded systems.  It is this second mystery that we address in this paper.

In considering \trans and \cis populations and their dynamics, we
assume  for simplicity that transitions are only optically-induced
in the forward, t$\rightarrow$c direction and that the
c$\rightarrow$t back-reaction is not optically, but only thermally
induced (at rate $1/\tau$), as is often the case
\cite{Eisenbach:78}; see also \cite{Statman:03}. We return to the case
of an optically stimulated back-reaction briefly at the end of the paper.  Given these optical
transitions, the spatial decay of the Poynting flux of a light beam,
$I(x)$, is governed by
 \bea \partial I/\partial x = - \gamma \Gamma n\s{t}(I) I \equiv - n\s{t}I / d \label{eq:spatial}
 \eea
where $n\s{t}$ is the local number fraction of \textit{trans} dye
molecules and depends on time $t$ and space $x$ (through $I$).  The
Beer length is $d=1/(\gamma \Gamma)$ that depends on the material
parameters $\gamma$ (proportional to the number density of
chromophores and the energy each t$\rightarrow$c transition absorbs
from the beam) and on $\Gamma$ which determines the t$\rightarrow$c
transition rate as we also see in the dynamical equation for
$n\s{t}$:
 \bea \partial n\s{t}/\partial t = -\Gamma I n\s{t} + n\s{c}/\tau \label{eq:trans_rate}.
  \eea
The \textit{cis} number fraction is $n\s{c} = 1 - n\s{t}$.  We
reduce intensity by the incident value to give $\cI(x) = I(x)/I_0$.
The combination $I\s{t} = 1/(\Gamma\tau)$ gives a characteristic
intensity, a material constant related to the \trans photo response.
If $I_0$ is reduced by $I\s{t}$, then $\alpha = I_0/I\s{t}$ is a
measure of how intense the incident beam is. The above  equations
then reduce to: \bea
  \partial \cI/\partial x = - n\s{t}\cI/d\; , \;\;  \partial n\s{t}/\partial t = -\left[(1 + \alpha \cI) n\s{t} - 1\right]/\tau \label{eq:reduced}
 \eea
At short times conversion has not yet proceeded and $n\s{t} = 1$.
Then the first of (\ref{eq:reduced}) easily integrates to  Beer's
Law $\cI(x, t\sim 0) = \e^{-x/d}$.

At long times (equilibrium) $\partial n\s{t}/\partial t = 0$ and
thus $n^{\infty}\s{t} = 1/(1+\alpha \cI)$.  For weak beams ($\alpha
\sim 0$) $n\s{t} \approx 1$ and again (\ref{eq:reduced}) gives
Beer's Law. At high intensities $\alpha \gg 1$ one has non-linear
absorption \cite{Statman:03}. Then $n\s{t} \sim 1/(\alpha\cI)$, at
least before depths such that the beam attenuates to $\cI \sim
1/\alpha$. Then in (\ref{eq:reduced}) one has $\partial \cI/\partial
x \approx -1/(\alpha d)$ and hence $\cI \simeq 1 - x/(\alpha d)$
(for $x\lesssim \alpha d$).  The intensity  profile is initially
linear until depths where $n\s{t}$ rises to being significantly
greater than $1/\alpha$ and absorption is then important.  The
surface layer is in effect photo-bleached and lets much light down
to depths $x > d$; see \cite{Corbett_PRL:07,Corbett_PRE:08} for a
detailed discussion, including of the role of optical backreaction.
We consider here for simplicity non mesogenic dyes and thus avoid
dye rotation rather than \textit{cis} formation as a route to
bleaching (see \cite{Yager:05} for this possibility). With the
equilibrium form of $n\s{t}$ inserted into the first part of
(\ref{eq:reduced}), we can integrate to give the full, non-linear,
equilibrium profile:
 \bea
 \ln[\cI(x)] + \alpha[ \cI(x) - 1]= - x/d \label{eq:profile} \; .
 \eea
 The long-time limit  ($t \gtrsim 5\tau$ in practice) of our dynamical profiles
 will display the linear rather than exponential forms until depths
 much greater than $d$.  The Mathematica defined function ProductLog gives \cite{Corbett_PRL:07,Corbett_PRE:08} the
solution: $\cI = \frac{1}{\alpha} {\rm
ProductLog}\left[\alpha\e^{(\alpha - x)}\right]$.

We are concerned here with the dynamics of the transition from
Beer's law initially to the above equilibrium profile for intense
beams.
 The absorbance $\cA = \ln\left[1/\cI(w,t)\right]$ is the usual measured quantity and reflects the absorber number in the optical path.
 Rearranging the first of eqns~
(\ref{eq:reduced}) to $\frac{1}{\cI}\frac{\partial \cI}{\partial x}
= -n\s{t}(x,t)/d$ and integrating $\int^{\cI}_1 d\cI$ and $\int_0^w
dx$, one obtains for all times and incident intensities (recall
$\cI(0, t) = 1$ for all $t$):
 \be \cA(w,t) = \frac{w}{d} \bar{n}\s{t} \equiv \frac{w}{d}\,
 \frac{1}{w} \!\int_0^w\!\!
dx \,n\s{t}(x,t) \label{eq:abs} \ee where $\bar{n}\s{t}$ is the mean
\trans number fraction through the sample.  In the non-linear limit,
$\bar{n}\s{t}$ is not independent of $w$ and hence $\bar{n}\s{t}/d$
is no longer a simple, material-dependent extinction coefficient.

The dynamics of $\cA$ is often observed and analysed assuming exponential behaviour:
 \be
\cA(w,t) \simeq \cA(w,\infty) + \left[\cA(w,0)- \cA(w,\infty)\right]
\e^{-t/(n\s{t}^{\infty}\tau)}.\nonumber
 \ee
Note that $\cA(w,0) = w/d$ exactly and that the non-linear limit of
$\cA(w,\infty)$ is given by the ProductLog solution to
eqn~(\ref{eq:profile}) for $\cI(w)$ .  However, this is not a solution to the differential eqns~(\ref{eq:reduced}) and becomes
a bad dynamical estimate for thick samples, $w\gg d$, that are only traversed by intense beams because of bleaching.  The characteristic time $\tau
n\s{t}^{\infty} =  \tau/(1+ \alpha \cI)$ is shorter than thermal
times when the driving $\alpha$ is large (intense beams), but
clearly $\cI(x)$ depends on position in thick samples and there is a
spectrum of times and the overall response is not exponential.
 \begin{figure}[!t]
\includegraphics[width=0.45\textwidth]{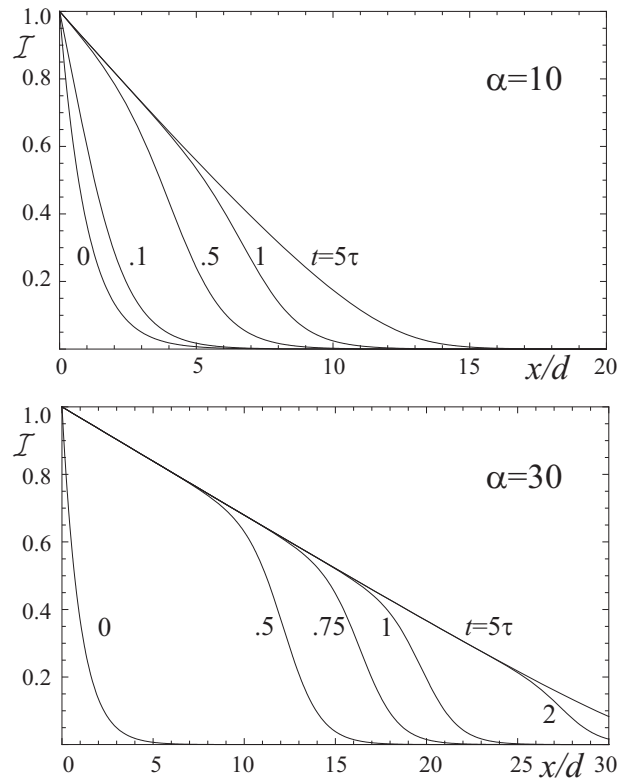}
\caption{Intensity against reduced depth for reduced incident
intensities $\alpha=10$ and 30 (reduced times
marked).}\label{fig:spatial}
\end{figure}
To solve the problem exactly, use $\cA(x,t) = - \ln(\cI)$ as the
variable.  Denote partial spatial and temporal derivatives of $X$ by
$X'$ and $\dot{X}$ respectively. Differentiating eqn~(\ref{eq:abs})
with respect to $t$ and using the second of eqns~(\ref{eq:reduced})
for $\dot{n}\s{t}$ under the integral yields:
 \begin{equation} \dot{\cA} = \int_0^x dx\left[1-(1+\alpha \cI)n\s{t}\right]/(\tau
 d) \nonumber \end{equation}
Now  use $n\s{t}/d = \cA'$, $\cA(0,t) = 0$, and $\cA' = -\cI'/\cI$:
\bea
\tau \dot{\cA} &=& x/d - \int_0^x dx(1+\alpha \cI)\cA' \label{eq:deriv-gen} \\
 &=& \frac{x}{d} -\cA + \alpha(\cI - 1) \equiv \frac{x}{d} -\cA +\alpha\left(\e^{-\cA}-1\right).
 \label{eq:deriv}
  \eea
 A final quadrature gives $\cA(x,t)$:
 \bea
  t/\tau =  \int^{\cA}_{x/d} \frac{d\cA}{x/d -\alpha -\cA + \alpha \e^{-\cA}} .\label{eq:conc} \eea
  The initial absorption $\cA(x,t=0) = x/d$ obtains from the vanishing of each side of (\ref{eq:conc}), and is Beer's Law.

Figure \ref{fig:spatial} first shows $\cI(x,t)$ $(\equiv
\e^{-\cA(x,t)})$ for reduced intensity $\alpha = I_0/I_c = 10$ as a
function of $x$ for a set of times $t$.
 Initially at $t=0$, the profile $\cI(x,0)$ is exponential, and at long times ($t=5\tau$)
the profile is essentially linear out to $x \sim \alpha d = 10 d$,
then decays exponentially. At intermediate times the profile first
saturates (bleaches) at small $x$, that is it approaches the initial
part of the equilibrium profile.  Then as the surface layers let
more light through, the profile deeper down also approaches the
non-Beer form. For even higher $\alpha=30$, higher incident
intensity, ultimate penetration is deeper and the approach to the
bleached state, as time advances, even sharper.  A front of
bleaching propagates through the sample.  Compare the two graphs at
a given $x=5d$, say.  The more intense case, $\alpha = 30$, sees a
much quicker achievement of the bleached state than the $\alpha =
10$ case, and over times considerably shorter than $\tau$, as the
approximate analysis suggested.  For either $\alpha$, a sample with
thickness $w=5d$, say, would see  an intensity of emerging light far
in excess of any Beer expectation, the rate build up of which we now
examine.

\begin{figure}[!t]
\includegraphics[width=0.45\textwidth]{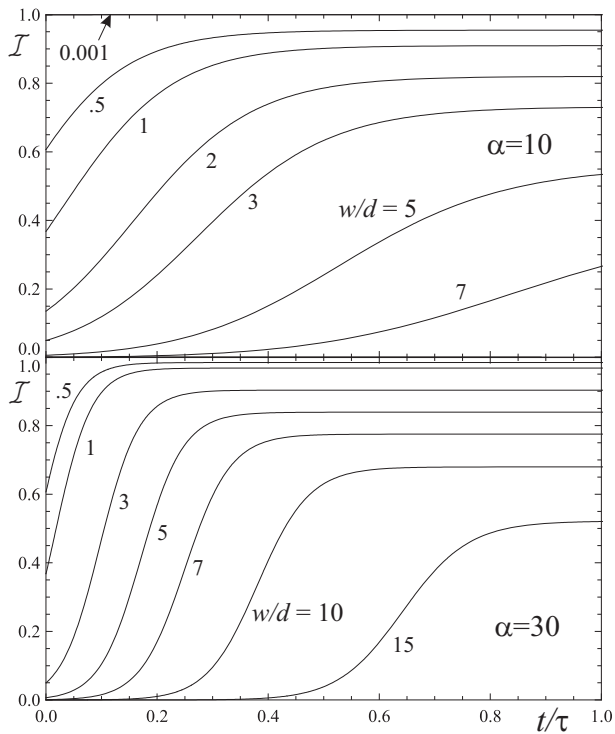}
\caption{Emergent reduced intensity $\cI = I(w)/I_0$ against time at
reduced incident intensity $\alpha=I_0/I_c = 10$ and $30$ for
various reduced
 sample thicknesses $w/d$.} \label{fig:temporal}
\end{figure}

Although these curves of Fig.~\ref{fig:spatial} are vital to
understanding the build up (and possible decay) of optically-induced
curvature in heavily dye-doped solid nematics, the intensity profile
is not in general directly observable.  It is more feasible to
measure the dynamics of the build-up of intensity of the light
emerging at the back face $x=w$ of a sample, $\cI(w,t)$.
 Figure \ref{fig:temporal} reveals $\cI(w,t)$ for various fixed thicknesses $w$, for two reduced incident intensities $\alpha$.
 The initial ($t=0$) value is
just that from Beer penetration, $\e^{-w/d}$.  Finally $\cI$ rises
to the long-time, bleached value $\cI(w,t\gg \tau)$ shown at  the
corresponding $x=w$ in Figs~\ref{fig:spatial}. The rise is naturally
slower for thicker samples.  Such dynamics of light penetration has
been seen by Serra and Terentjev \cite{Serra:08}.

\begin{figure}[!t]
\includegraphics[width=0.46\textwidth]{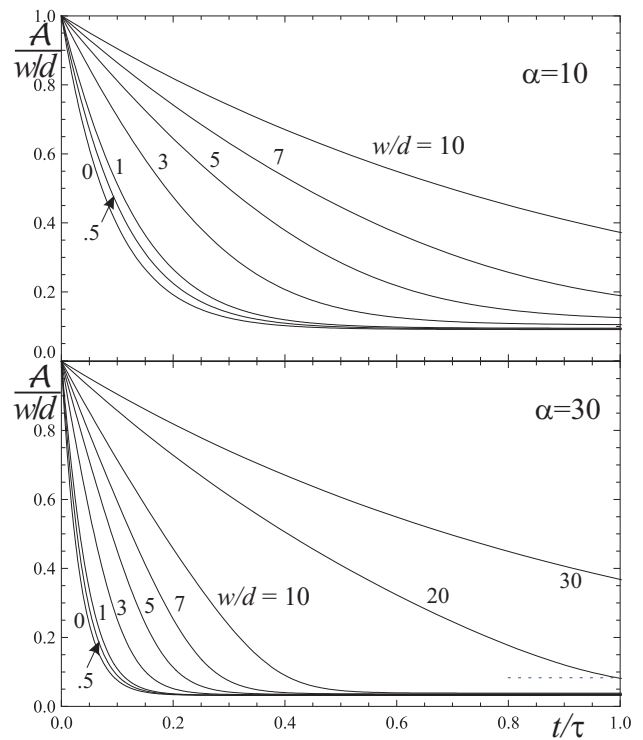}
\caption{Non-linear extinction $\cA(w,t)/(w/d)$ against time at
reduced incident intensity $\alpha=I_0/I_c = 10$ and $30$ for
various reduced
 sample thicknesses $w/d$.  The dotted line is the asymptote for the $w/d=30$ dynamics.} \label{fig:extinct}
\end{figure}

Thick ($w>d$) beams give initially exponentially small emergent
fluxes that give way to large fluxes as the linear profile is set
up.  Such data are accommodated in the usual logarithmic way,
here by dividing the absorbance by the thickness which is not a
material-dependent constant (an extinction) in this non-linear
limit. Figure \ref{fig:extinct} shows  $\cA(w,t)/(w/d)$ for various
fixed thicknesses $w$, for incident intensities $\alpha=10,\,30$.

The initial increase of intensity away from the Beer value in
Fig.~\ref{fig:temporal} is $\dot{\cI}(w,t=0) = -
\dot{\cA}(w,0)\e^{-\cA(w,0)} \equiv -\dot{\cA}(w,0)\e^{-w/d}$
whereupon using (\ref{eq:deriv}) for $\dot{\cA}$ at $t=0$ where
$\cA=w/d$, one obtains for initial slope
 \bea \tau\dot{\cI}(w,t=0) = \alpha\left(1-\e^{-w/d}\right)\e^{-w/d}\label{eq:initial_deriv}. \eea
 For intense beams ($\alpha \gg 1$) on thick ($w \sim \alpha d$) samples, the initial intensities at the back face
 on irradiating the front face are small, $\cI(w,0) = \e^{-w/d}$. The rates of increase are also initially very
 small, $\dot{\cI} \sim \alpha \e^{-\alpha}/\tau$, but then rise sharply with time, see the example with $\alpha = 30$.

 Two other processes are sometimes important in non-linear
 absorption, those of host absorption and of \cis absorption, that
 is optically-stimulated back reaction.  Both are discussed in
 detail in the non-linear static case \cite{Corbett_PRE:08}.
 Moderate host absorption can be successfully divided out and plays
 a limited role.  Back reaction occurs if the t$\rightarrow$c and
 c$\rightarrow$t absorption lines begin to overlap.  This is
 sometimes the case and has the effect of reducing deep penetration \cite{Statman:03}.

 In
 eqn~(\ref{eq:spatial}) host absorption gives another term, $-I/d\s{h}$, while \cis depletion of the beam gives
 $-\gamma\s{c}\Gamma\s{c} n\s{c}(I) I$.
 Optical back reaction also effects the dynamics.  An additional
 term $+\Gamma\s{c} I n\s{c}$ from the decay c$\rightarrow$t acts in eqn~(\ref{eq:trans_rate}) to
 replenish the \trans population.  There is a characteristic
 intensity $I\s{c} = 1/(\Gamma\s{c}\tau)$ analogous to that of the
 \trans species, and thus another measure $\beta = I_0/I\s{c}$ of
 the incident beam intensity.  Now the coupled, non-linear partial
 differential equations~(\ref{eq:reduced}) become:
\bea
\cA' &=&  - \frac{\partial \cI/\partial x}{\cI} =  \left( \frac{1}{d} - \frac{1}{d\s{c}}\right)n\s{t}  +  \frac{1}{d\s{c}} +  \frac{1}{d\s{h}}\label{eq:host-abs}\\
 \tau \dot{n}\s{t} &=& (1+\beta\cI) - [1 + (\alpha +
\beta) \cI] n\s{t} .\label{eq:host-cis}
 \eea
 The changes, even if only one of these influences is introduced at
 a time, add difficulty to the solution of the equations.  Proceeding as before, one
 differentiates eqn~(\ref{eq:host-abs}) with respect to time and
 uses (\ref{eq:host-cis}) to eliminate $\dot{n}\s{t}$.  Where
 $n\s{t}$ appears, it can be eliminated in favour of $\cA'$ by using
 eqn~(\ref{eq:host-abs}) again.  Terms do not quite all integrate
 totally as in going from (\ref{eq:deriv-gen}) to (\ref{eq:deriv})
 because eqn~(\ref{eq:host-abs}) has additional new constant terms.
 \begin{figure}[!t]
\includegraphics[width=0.45\textwidth]{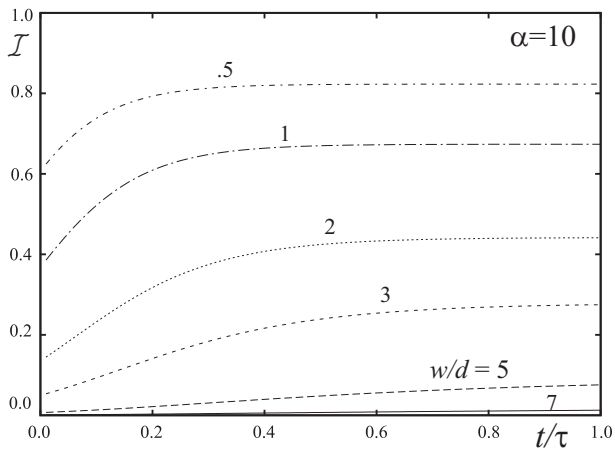}
\caption{Emergent reduced light intensity $\cI(w,t)$ against time
for various reduced sample thicknesses $w/d$. Reduced incident
intensity is $\alpha=I_0/I_c = 10$ with a relatively high degree of
reduced optical back reaction $\beta = 2$.}
\label{fig:optical-back}
\end{figure}
 Spatially integrating gives:
 \bea \tau\dot{\cA} = &-&\cA + x/d\s{eff} + (\alpha + \beta)(\cI - 1)
 + \nonumber \\
&+&  \left[\alpha\left(\frac{1}{d\s{c}} + \frac{1}{d\s{h}}\right)
 +\beta/d\s{eff}\right]\int^x_0 dx \cI \label{eq:gen-dyn}
 \eea
where $1/d\s{eff} = 1/d + 1/d\s{h}$ is an effective absorption
length arising from the simple combination of dye and host
absorptions in the Beer limit.  It is interesting that although
$\beta$ (and thus $1/d\s{c}$) and $1/d\s{h}$ enter
eqns~(\ref{eq:host-abs}) and (\ref{eq:host-cis}) in entirely
different ways, their effects can be scaled on to each other in the
resultant equation for dynamical non-linear absorption
(\ref{eq:gen-dyn}).  Remember however that $d\s{eff}$ depends on
$d\s{h}$ and so lengths are not quite equivalently effected by this
interchange of sources of extra absorption.
 We show just the effect
of adding in optical back-reaction into the dynamical equations,
that is $1/d\s{h} = 0$ and lengths are still reduced by $d$. For
$\alpha = 10$ and, say, $\beta = 0.2$, the $\cI(w,t)-t$ curves are
indistinguishable from those in upper Fig.~\ref{fig:temporal} which
have $\beta = 0$. Figure~\ref{fig:optical-back} shows the behaviour
for $\beta = 2$ which should be compared with upper
Fig.~\ref{fig:temporal}.
  The starting values are of course the
same; the final values are lower because $\beta$ acts to lower the
final penetration.  The upward curvature at short times is lost.
Thicker samples are much more drastically affected.

We have shown that the dynamics of penetration of intense light
beams into heavily absorbing media is complex and strongly dependent
upon incident intensity.  A practical experiment to observe this
effect via the emergent light from the sample is proposed.  There is
evidence from large amplitude photo-mechanics that non-linear
effects are important in practice.

\vspace{.2cm} \noindent \textit{Acknowledgements.}  We are grateful
to Kees Bastiaansen, Dick Broer, Francesca Serra and Eugene Terentjev for useful discussions, and to David Statman for guidance on non-linear statics and dynamics.


\end{document}